\documentclass[aps, pra, floatfix, reprint]{revtex4-1}

\usepackage{graphicx}
\usepackage{amsmath}
\usepackage{color}
\usepackage{amssymb}
\usepackage{textcomp}
\usepackage{bm}
\usepackage{hyperref}

\hypersetup{
	colorlinks = true, 
	urlcolor   = blue, 
	linkcolor  = blue, 
	citecolor  = blue,   
}

\begin{document}
	\title{Ferroelectrics Manipulate Magnetic Bloch Skyrmions in a Composite Bilayer}
	\author{Zidong Wang}
	\email{Zidong.Wang@auckland.ac.nz}
	\author{Malcolm J. Grimson}
	\email{m.grimson@auckland.ac.nz}
	\affiliation{Department of Physics, University of Auckland, Private Bag 92019, Auckland, New Zealand}
	\date{\today}

	\begin{abstract}
		Theoretical investigation demonstrates that the composite bilayer (i.e., chiral-magnetic/ferroelectric bilayer) offers the possibility of electric-induced magnetic Skyrmions [Phys. Rev. B \textbf{94}, 014311 (2016)]. In this Article, we propose a micromagnetic model to physically manipulate magnetic Bloch Skyrmions propagating in a chiral-magnetic thin film with a polarized ferroelectric essential to drive the system through the converse magnetoelectric effect. Effects caused by different velocities of the propagation, sizes of the thin film, and strength of the magnetoelectric couplings strongly impact on quality and quantity of the magnetic Skyrmions.\\
		\\
		\noindent E-print: \url{http://arxiv.org/abs/1604.08780}
	\end{abstract}
	
\maketitle

\section{Introduction}
\label{Introduction}

Magnetic Bloch Skyrmions behave as stable particle-like spin textures in the chiral-magnetic crystals \cite{N.442.797}, such as \textit{B20} compound metallic $\text{MnSi}$ \cite{S.323.915}, $\text{FeGe}$ \cite{NM.10.106}, $\text{Fe}_{1-x}\text{Co}_x\text{Si}$ \cite{N.465.901}, $ \text{MnGe} $ and $\text{Mn}_{1-x}\text{Fe}_x\text{Ge}$ \cite{NN.8.723}, and multiferroic $\text{Cu}_2\text{OSeO}_3$ \cite{S.336.198}. These materials have no inversion symmetry that can allow the emergence of magnetic Bloch Skyrmions, due to their non-centrosymmetric lattice structures \cite{Nanotech.26.225701}. In micromagnetics, this phenomenon is caused by two components: the nearest-neighbor (symmetric exchange) interaction and the inherent Dzyaloshinskii-Moriya (asymmetric exchange) interaction \cite{NN.8.899}. The competition between them stabilizes the helicity of magnetic Skyrmions \cite{PRB.88.195137}. Mathematically, the Dzyaloshinskii-Moriya interaction $\mathcal{H}_{\text{dmi}}$ is the contribution of a non-linear exchange interaction between two neighboring magnetic spins, $\bm{S}_1$ and $\bm{S}_2$ \cite{JPCS.4.241,PR.120.91}, written as 
\begin{equation}
	\mathcal{H}_{\text{dmi}} = \bm{D} \cdot [\bm{S}_1 \times \bm{S}_2],
	\label{Eq.DMI}
\end{equation}
where $\bm{D}$ is an oriented vector, which indicates the constrained helicity to the symmetric state. The nearest-neighbor interaction $\mathcal{H}_{\text{int}}$ commonly exists in ferromagnets as a linear exchange interaction,
\begin{equation}
	\mathcal{H}_{\text{int}} = J [\bm{S}_1 \cdot \bm{S}_2],
	\label{Eq.Int}
\end{equation} 
where $J$ is the termed exchange coupling. Magnetic Skyrmion holds great potential for applications in spintronic memory devices, due to their self-protection behavior.

So far, magnetic Bloch Skyrmions have been observed in insulating multiferroics, i.e., $\text{Cu}_2\text{OSeO}_3$ \cite{S.336.198}. This multiferroism offers an opportunity to generate magnetic Skyrmions by electric polarization \cite{PRL.113.107203}. It is due to the converse magnetoelectric effect, which is the phenomenon of inducing magnetization by applying an external electric field \cite{PRB.87.100402}. Unfortunately, the multiferroic insulators require a low transition temperature, and have a limited magnetic response, which is adverse for applications \cite{PRB.78.094416}. But composite multiferroics, which are an artificially synthesized heterostructure of ferromagnetic and ferroelectric order, have a remarkable magnetoelectric coupling due to the microscopic mechanism of the strain-stress effect \cite{PRB.50.6082}. This coupling describes the influence of electric polarization on the magnetization at interface \cite{APL.106.262902}.

A previous investigation has discussed the magnetic Bloch Skyrmions induced by an electric driving field in a composite chiral-magnetic (CM) and ferroelectric (FE) bilayer \cite{PRB.94.014311}. In this Article, we demonstrate a micromagnetic model in Sec.~\ref{Model}, for generating and manipulating the magnetic Bloch Skyrmions in a CM thin film, which is driven by a piece of mobile and polarized FE thin film. Both of films are glued by a strong magnetoelectric coupling. The dynamical behaviors in the CM layer and the dynamics of the FE layer are described in Sec.~\ref{Method}. Results in Sec.~\ref{Results} show the creation and propagation of magnetic Bloch Skyrmions, including effects of the propagation velocity in Sec.~\ref{Velocities}, the size of thin film in Sec~\ref{Sizes}, and the strength of magnetoelectric coupling in Sec.~\ref{Couplings}. The paper concludes with a discussion in Sec.~\ref{Conclusion}.

\section{Model}
\label{Model}

\begin{figure}
	\includegraphics[width = 250px, trim = 0 490 0 400, clip]{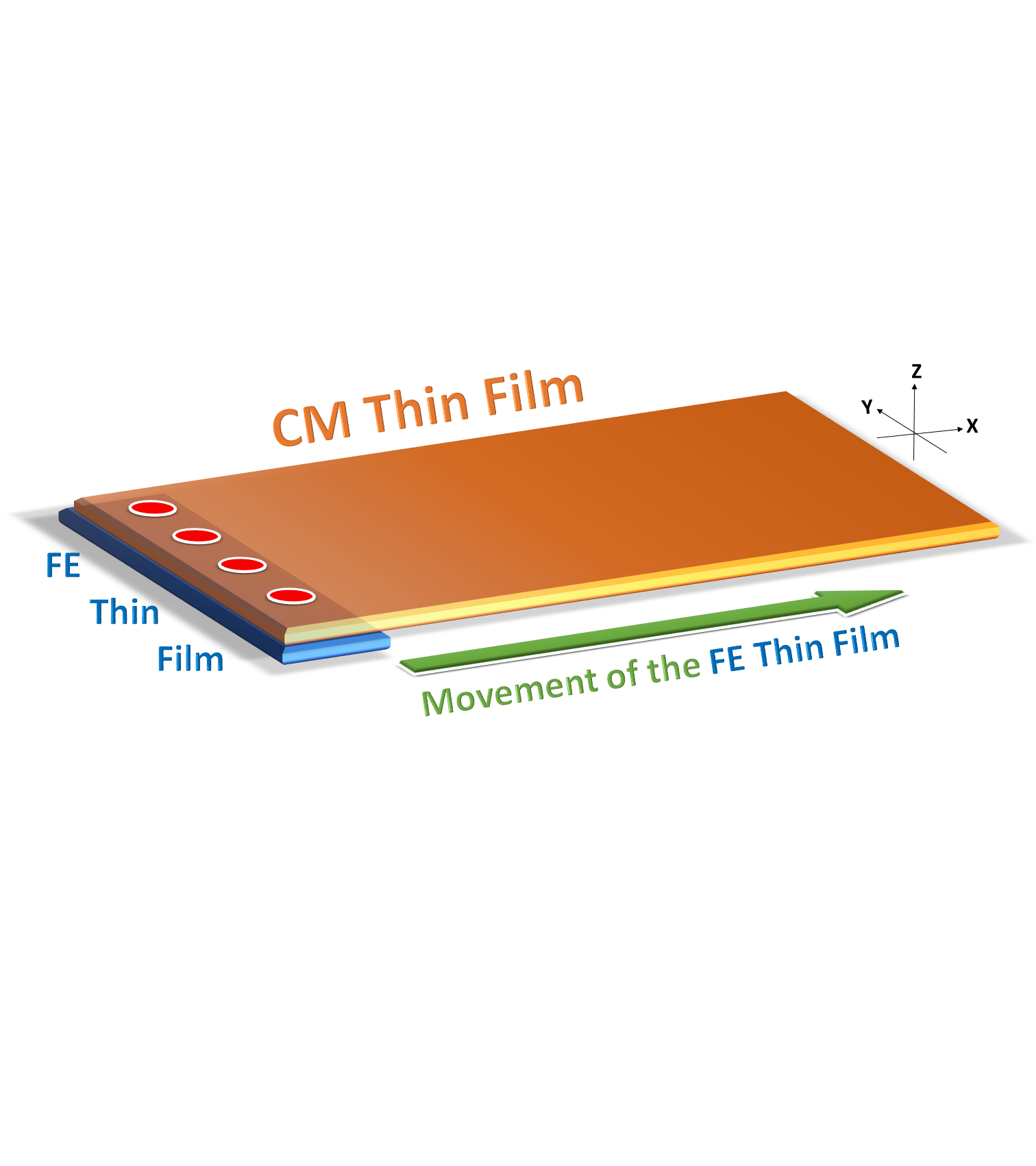}
	\caption{ \textbf{Schematic of the CM/FE heterostructure bilayer.} The top layer represents the CM thin film, which can construct magnetic Bloch Skyrmions, as shown in the red circles; The smaller FE thin film is movable and coupled with the CM thin film. See \textbf{Movie 1} in the Supplemental Material \cite{Supplementary}. }
	\label{Fig.1}
\end{figure}

\textbf{Figure~\ref{Fig.1}} illustrates the model of a composite bilayer, which consists a CM thin film at top, and a mobile FE thin film attached at bottom. The CM film can hold the magnetic Bloch Skyrmions. The FE film has a smaller size, and can be physically driven by the technology of microelectromechanical systems under the CM film. The combination between them induces the converse magnetoelectric effect. The animation of this dynamics is shown in \textbf{Movie 1} in the Supplemental Material \cite{Supplementary}.

The magnetic component of the system is described by the classical Heisenberg model in a two-dimensional rectangular lattice. The magnetic spin is represented by $\bm{S}_{i,j} = (S_{i,j}^x , S_{i,j}^y , S_{i,j}^z)$, which is a normalized vector, i.e., $\| \bm{S}_{i,j} \| = 1$, and $i,j \in [1,2,3,...,N]$ characterizes the location of each spin in the film. The Hamiltonian $\mathcal{H}$ is defined by
\begin{eqnarray}
	\mathcal{H} & = & -J \sum_{i,j} [ \bm{S}_{i,j} \cdot ( \bm{S}_{i+1,j} + \bm{S}_{i,j+1}) ]\nonumber\\
                &   & -D \sum_{i,j} [ \bm{S}_{i,j} \times \bm{S}_{i+1,j} \cdot \hat{x} + \bm{S}_{i,j} \times \bm{S}_{i,j+1} \cdot \hat{y} ]\nonumber\\
                &   & -K^z \sum_{i,j} (S_{i,j}^z)^2\nonumber\\
                &   & -g \sum_{\tilde{i},\tilde{j}} (S_{\tilde{i},\tilde{j}}^z P).
	\label{Eq.Ham}
\end{eqnarray}
The first term represents the nearest-neighbor exchange interaction, and $J^* = J / k_B T$ is the dimensionless exchange coupling coefficient. The second term represents the two-dimensional Dzyaloshinskii-Moriya interaction \cite{NN.8.899} , and $D^* = D / k_B T$ is the dimensionless Dzyaloshinskii-Moriya coefficient, and $\hat{x}$ and $\hat{y}$ are the unit vectors of the \textit{x}- and \textit{y}-axes, respectively. The third term represents the magnetic anisotropy, and $K^* = K / k_B T$ represents the dimensionless uniaxial anisotropic coefficient in the \textit{z}-direction. The last term represents the magnetoelectric coupling, which is generally described as a linear spin-dipole interaction \cite{PRB.90.054423}, where $g^* = g / k_B T$ is the dimensionless strength of the magnetoelectric coupling. The analytic expression of the magnetoelectric coupling can be linear or non-linear, particularly with respect to the thermal effect \cite{PRB.92.134424}. A non-linear expression has not been studied here, for simplicity and due to their minor effects in the micromagnetic modeling. Note that, the magnetoelectric coupling was discussed by Spaldin \textit{et al.} \cite{PRB.93.195167}. The strength of coupling is, however, unknown. Hence, we only use the low-energy excitations between the CM and FE layers. So we restrict ourselves to the linear expression of the magnetoelectric interaction. The coupling sites of magnetic spins, $\tilde{i} , \tilde{j}$, to the FE layer are variable. This is a result of a polarization pulse propagating through the CM layer. Beach \textit{et al.} have physically built a similar model in metallic ferromagnets with electric current-driven dynamics \cite{NM.15.501}.

\section{Method}
\label{Method}

The dynamics of magnetic spins in the CM layer has been studied by the Landau-Lifshitz equation \cite{JPD.48.30500}, which numerically solves the rotation of a magnetic spin in response to its torques \cite{JAP.118.124109},
\begin{equation}
	\frac{\partial \bm{S}_{i,j}}{\partial t} = -\gamma [\bm{S}_{i,j} \times \bm{H}_{i,j}^{\text{eff}}] - \lambda [\bm{S}_{i,j} \times (\bm{S}_{i,j} \times \bm{H}_{i,j}^{\text{eff}})],
	\label{Eq.LL}
\end{equation}
where $\gamma$ is the gyromagnetic ratio which relates the spin to its angular momentum, and $\lambda$ is the phenomenological damping term. $\bm{H}_{i,j}^{\text{eff}}$ is the effective field acting on each magnetic spin, which is the functional derivative of the system Hamiltonian [Eq. (\ref{Eq.Ham})] with respect to the magnitudes of the magnetic spin in each direction \cite{JAP.119.124105}, as $\bm{H}_{i,j}^{\text{eff}}  = - \dfrac{\delta \mathcal{H}}{\delta \bm{S}_{i,j}}$. This Landau-Lifshitz equation is solved by a fourth-order Range-Kutta method with a dimensionless time increment $\Delta t^* = 0.0001$ of in all simulations.

The CM layer is large and stationary, the polarized FE layer is under the CM layer has a much smaller size. Thus FE layer is transversely traveling along the CM layer with a certain velocity. This technology refers to the microelectromechanical systems. In simulations, the electric dipoles in the FE layer are coupled locally with magnetic spins in the CM layer. So the FE layer moves at a certain rate which characterizes the propagation velocity of Skyrmions in the CM layer. \textbf{Movie 1} shows the animation of this dynamics in the Supplemental Material \cite{Supplementary}.

\section{Results}
\label{Results}

To investigate the dynamical manipulation of magnetic Bloch Skyrmions, we implement a dimensionless parameter set: $J^* = 1$, $D^* = 1$, $K^* = 0.1$, $g^* = 0.5$, $\gamma^* = 1$, and $\lambda^* = 0.1$. Note that \textquotedblleft $*$\textquotedblright~characterizes dimensionless quantity. The CM layer with $N_{\text{CM}} = 20 \times 100$ magnetic spins, and $N_{\text{FE}} = 20 \times 20$ electric dipoles in the FE layer are used. Free boundary conditions and a random initial state are applied. The propagation step-time is measured as the non-dimensional time period stopped on each position, like an intermittent pulse, with a magnitude of $T^* = 20/\text{step}$.

\begin{figure}
	\includegraphics[width = 250px]{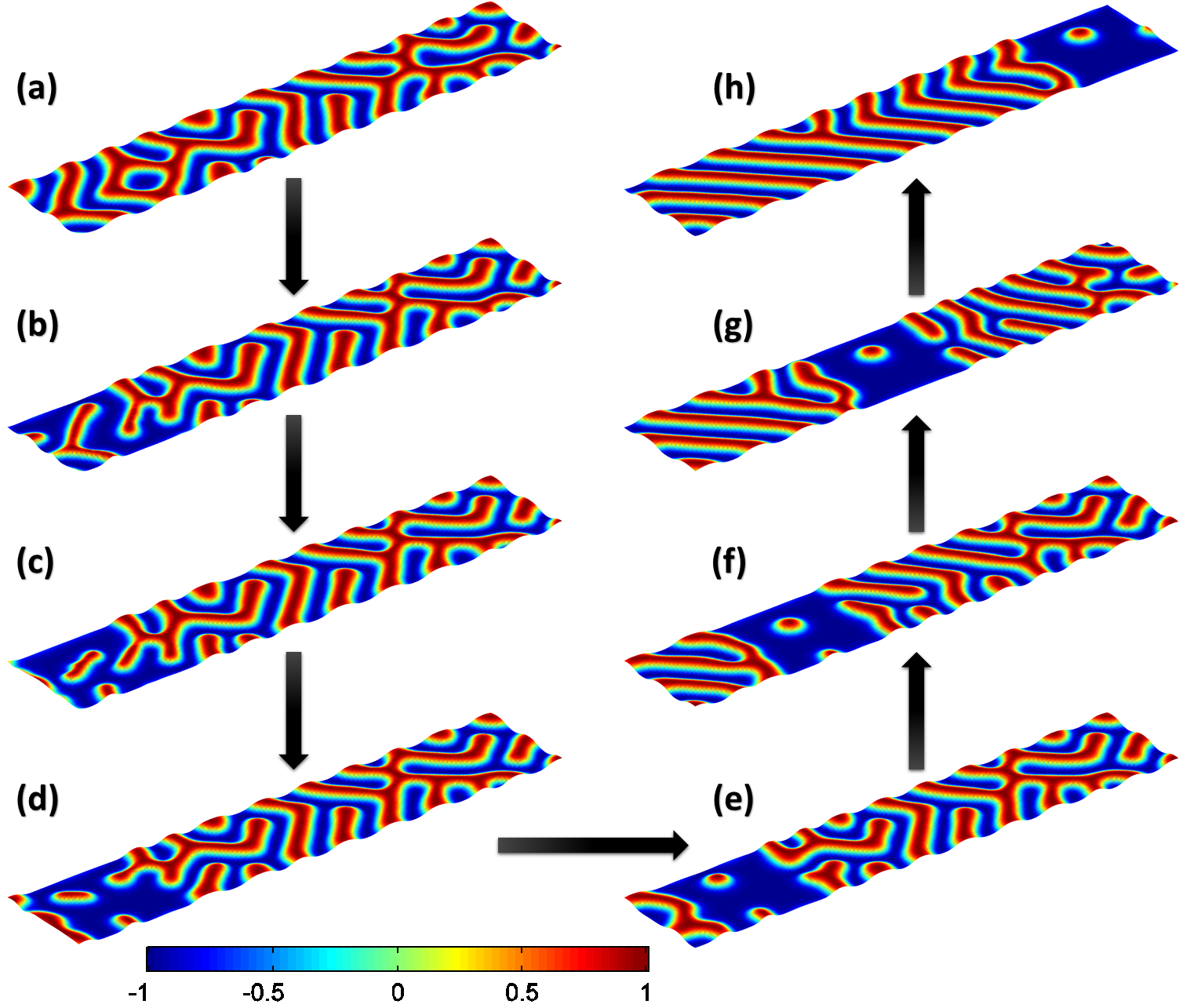}
	\caption{ \textbf{Generation and propagation of a magnetic Bloch Skyrmion in the CM layer.} \textbf{(a)} An randomly helimagnetic state at start. \textbf{(b)$\rightarrow$(c)$ \rightarrow$(d)$\rightarrow$(e)} images show details of a Skyrmion generation in the CM film. \textbf{(e)$\rightarrow$(f)$\rightarrow$(g)$\rightarrow$(h)} images show details of the Skyrmion propagating through the CM film. The color scale represents the magnitude of the local \textit{z}-componential magnetization. See \textbf{Movie 2} in the Supplemental Material \cite{Supplementary}. }
	\label{Fig.2}
\end{figure}

\textbf{Figure~\ref{Fig.2}} summarizes a time evolution that generates a magnetic Bloch Skyrmion and manipulates it propagating along the CM layer. The CM layer is contacted with a polarized FE thin film, which starts from the left, then moves to right as shown in \textbf{Fig.~\ref{Fig.1}}. Initially, the magnetic domain walls are randomly located without any external energy addition on the system in \textbf{Fig.~\ref{Fig.2}(a)}. \textbf{Figures~\ref{Fig.2}(b)$\rightarrow$(c)$\rightarrow$(d)$\rightarrow$(e)} show the generation of a Skyrmion from natural alignment. Subsequently, this Skyrmion propagates though the CM layer, as shown in \textbf{Figs.~\ref{Fig.2}(e)$\rightarrow$(f)$\rightarrow$(g)$\rightarrow$(h)}. Eventually, this Skyrmion stops at the right-hand side of thin film [\textbf{Fig.~\ref{Fig.2}(h)}]. The fully dynamical process is shown in \textbf{Movie 2} in the Supplemental Material \cite{Supplementary}.

As seen in \textbf{Figs.~\ref{Fig.2}(d)$\rightarrow$(e)$\rightarrow$(f)$\rightarrow$(g)}, another partial Skyrmion at the edge been devoured, due to the free boundary condition. This can be avoided by using periodic boundary conditions. This will be studied later in Sec.~\ref{Sizes}: \textit{Sizes of Thin Film}. Additionally, passage of the FE film leaves a spin spiral alignment in the CM layer. This is due to the existence of a finite Dzyaloshinskii-Moriya interaction in equilibrium.

\subsection{Velocity of Propagation}
\label{Velocities}
We next discuss the effects due to the propagation velocity to the Skyrmions. Since the transverse travel of FE thin film can be manually controlled, Skyrmions are tracking this FE layer with the velocity with a short relaxation time. In this case, a large propagation velocity represents short time period of the FE layer to stay in one position (or one step), i.e., small traveling step-time.

\begin{figure}
	\includegraphics[width = 250px, trim= 0 20 0 0, clip]{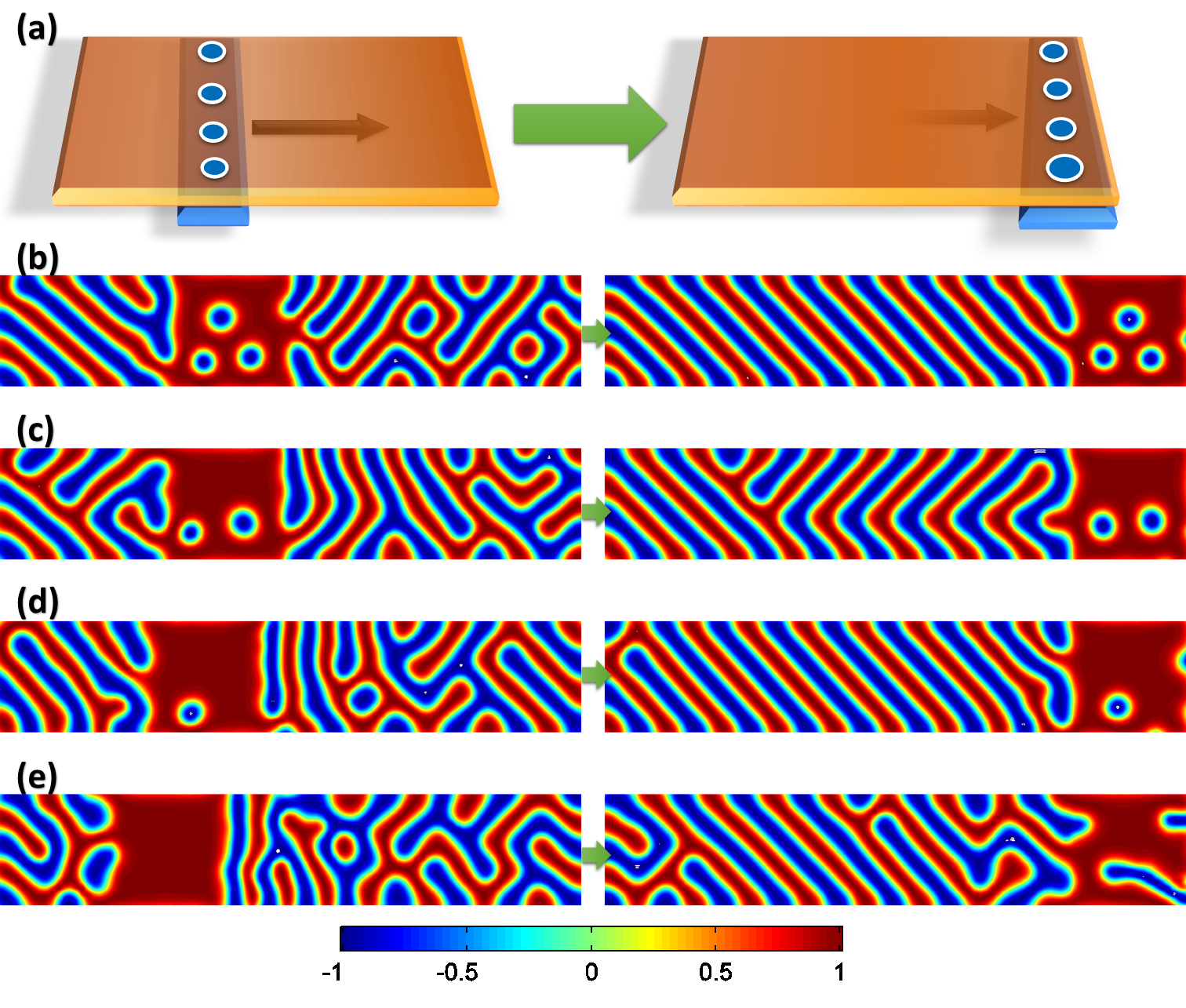}
	\caption{ \textbf{Propagating Skyrmions with different traveling step-time, $T^*$.} \textbf{(a)} Schematic of two issues: traveling and finishing. \textbf{(b)} $T^* = 30/\text{step}$, \textbf{(c)} $T^* = 20/\text{step}$, \textbf{(d)} $T^* = 10/\text{step}$, and \textbf{(e)} $T^* = 5/\text{step}$. The color scale represents the magnitude of the local \textit{z}-componential magnetization. See \textbf{Movie 3} in the Supplemental Material \cite{Supplementary}. }
	\label{Fig.3}
\end{figure}

\textbf{Figure~\ref{Fig.3}(a)} shows two schematics demonstrating the general issues of traveling (left) and finishing (right) in the following results. In \textbf{Fig.~\ref{Fig.3}(b)}, a slow propagation velocity with the traveling step-time $T^* = 30/\text{step}$. Three Skyrmions have been generated and propagated. As we drop the step-time to $T^* = 20/\text{step}$ in \textbf{Fig.~\ref{Fig.3}(c)}, the number of Skyrmions reduces to two, but they are have similar size as in \textbf{Fig.~\ref{Fig.3}(b)}. Subsequently, only one Skyrmion survived in \textbf{Fig.~\ref{Fig.3}(d)} with $T^* = 10/\text{step}$. If a step-time shorter than this value is used, it is too fast to create and propagate Skyrmions in the CM layer. This has been shown in \textbf{Fig.~\ref{Fig.3}(e)} with $T^* = 5/\text{step}$. The dynamical processes and the velocities comparison are shown in \textbf{Movie 3} in the Supplemental Material \cite{Supplementary}.

\subsection{Size of Thin Film}
\label{Sizes}
\begin{figure}
	\includegraphics[width = 250px, trim= 0 180 0 0, clip]{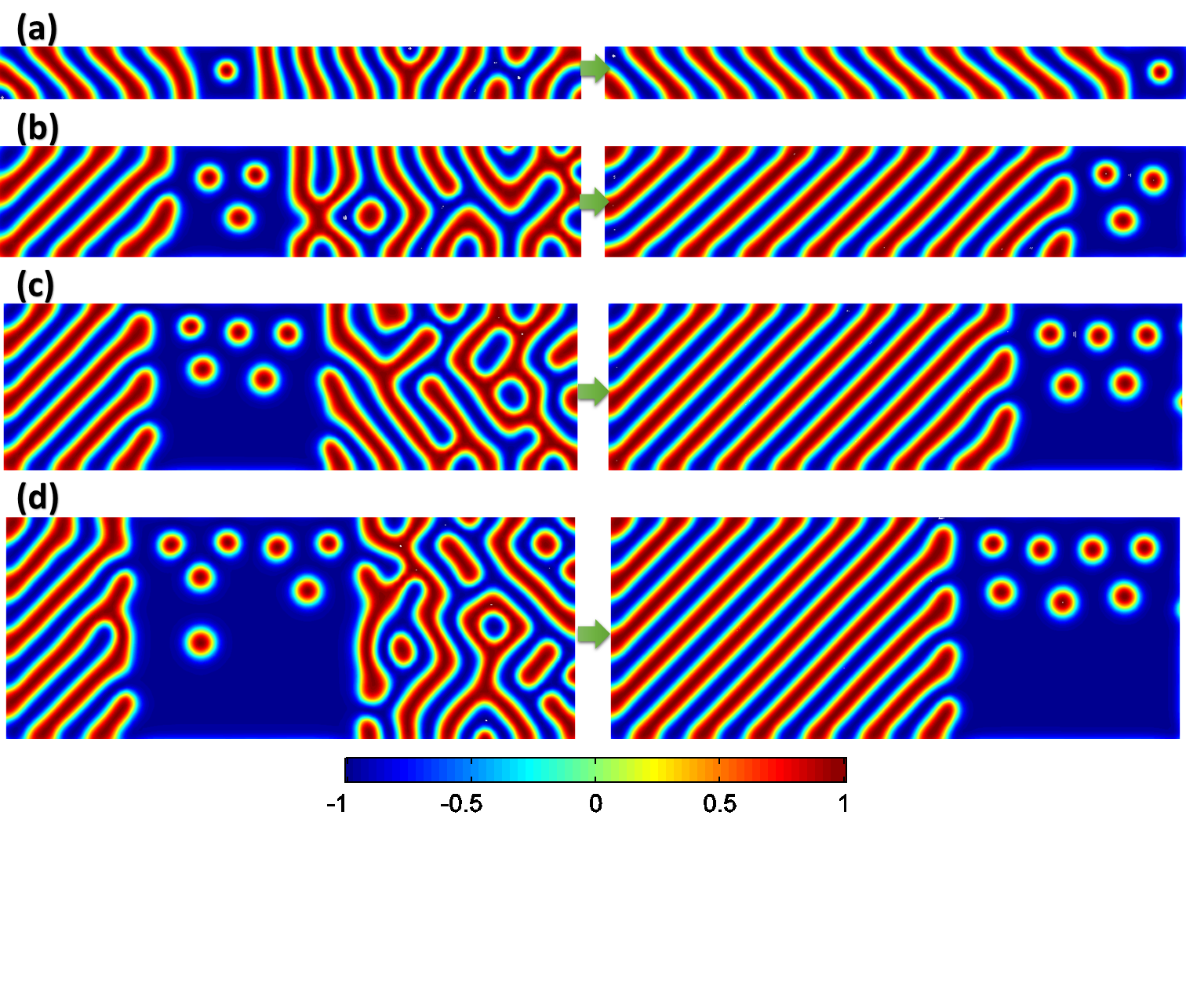}
	\caption{ \textbf{Effects by different sizes of the CM thin film, $N_{\text{CM}}$.} \textbf{(a)} $N_{\text{CM}} = 10 \times 100$, \textbf{(b)} $N_{\text{CM}} = 20 \times 100$, \textbf{(c)} $N_{\text{CM}} = 30 \times 100$, and \textbf{(d)} $N_{\text{CM}} = 40 \times 100$. Each subplot shows two issues as shown in \textbf{Fig.~\ref{Fig.3}(a)}. The color scale represents the magnitude of the local \textit{z}-componential magnetization. See \textbf{Movie 4} in the Supplemental Material \cite{Supplementary}. }
	\label{Fig.4}
\end{figure}

Larger size of the thin film offers more space to allow more Skyrmions. \textbf{Figure~\ref{Fig.4}} exemplifies four cases for different sizes of the composite bilayer. \textbf{Figure~\ref{Fig.4}(a)} shows the CM layer contains $N_{\text{CM}} = 10 \times 100$ magnetic spins, which generates one Skyrmion. Subsequently, a larger layer with $N_{\text{CM}} = 20 \times 100$ magnetic spins shows three Skyrmions in \textbf{Fig.~\ref{Fig.4}(b)}. \textbf{Figure~\ref{Fig.4}(c)} shows five Skyrmions with a layer size of $N_{\text{CM}} = 30 \times 100$ magnetic spins, and \textbf{Fig.~\ref{Fig.4}(d)} shows seven Skyrmions with a layer size of $N_{\text{CM}} = 40 \times 100$ magnetic spins. Consequently, the quantity of Skyrmions increased as the film size increases, but the quality of Skyrmions is the same. \textbf{Movie 4} in the Supplemental Material animates these cases \cite{Supplementary}.

Interestingly, Skyrmions are found to collect together near the top of thin film in \textbf{Figs.~\ref{Fig.4}(b)}, \textbf{(c)} and \textbf{(d)}. This occurs due to the helimagnetically ordered structure which points to the upper-right corner, and generates Skyrmions in this direction. Remember that the left hand side in each image is the structure after FE film has passed. This behavior is also observed in \textbf{Figs.~\ref{Fig.3}(c)} and \textbf{(d)}. 

In another simulation, we replace the free boundaries by a periodic boundary condition, which linked the spins at the top and the bottom. Now the dynamics shows Skyrmions on a race track moving longitudinally. Different to the behavior shown from the system with free boundary condition. See \textbf{Movie 6} in the Supplemental Material \cite{Supplementary}.

\subsection{Strength of Magnetoelectric Coupling}
\label{Couplings}
\begin{figure}
	\includegraphics[width=250px, trim= 0 50 0 0, clip]{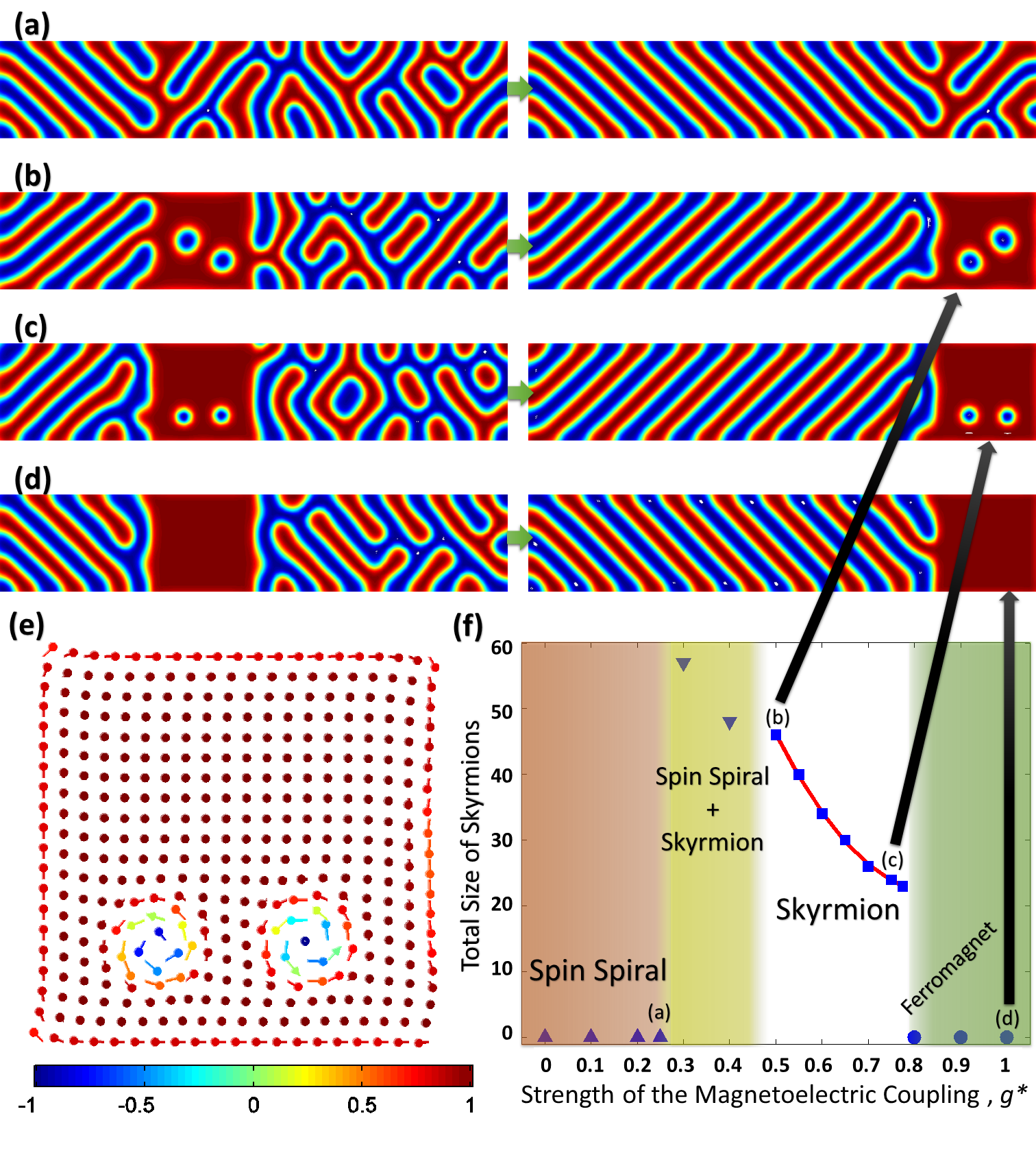}
	\caption{ \textbf{Effects by different strength of the magnetoelectric couplings, $g^*$.} \textbf{(a)} $g^* = 0.25$, \textbf{(b)} $g^* = 0.5$, \textbf{(c)} $g^* = 0.75$ and \textbf{(d)} $g^* = 1$.  Each subplot shows two issues as shown in \textbf{Fig.~\ref{Fig.3}(a)}. See \textbf{Movie 5} in the Supplemental Material \cite{Supplementary}. \textbf{(e)} A spin-plot image shows the Skyrmion part from right image in \textbf{(c)}. The color scale represents the magnitude of the local \textit{z}-componential magnetization. \textbf{(f)} The total size of Skyrmions versus the strength of magnetoelectric coupling, determined from the finishing image of each dynamics. The curve through the points is a guide to the eye. See \textbf{Fig. 1} in the Supplemental Material \cite{Supplementary}. }
	\label{Fig.5}
\end{figure}

The electric-induced magnetic Bloch Skyrmions result from the magnetoelectric coupling between the electric dipoles to the magnetic spins. It is noteworthy that the strength of magnetoelectric coupling plays an important role in mediating the energy transfer to sustain the magnetic Skyrmions \cite{PRB.94.014311}. Therefore, we examined different magnetoelectric coupling strength in \textbf{Fig.~\ref{Fig.5}}. Such as, $g^* = 0.25$ in \textbf{Fig.~\ref{Fig.5}(a)}, $g^* = 0.5$ in \textbf{Fig.~\ref{Fig.5}(b)}, $g^* = 0.75$ in \textbf{Fig.~\ref{Fig.5}(c)} and $g^* = 1$ in \textbf{Fig.~\ref{Fig.5}(d)}. Firstly, \textbf{Fig.~\ref{Fig.5}(a)} shows an insufficient strength of the magnetoelectric coupling to generate Skyrmions. The magnetic domain shows a spin spiral alignment. With increased strength of the magnetoelectric coupling in \textbf{Figs.~\ref{Fig.5}(b)} and \textbf{(c)}, both of them generate and propagate two Skyrmions in the CM layer. \textbf{Figure.~\ref{Fig.5}(b)} has larger size of Skyrmions than in \textbf{Fig.~\ref{Fig.5}(c)}, since the magnetoelectric interaction in the former is weak compared to its Dzyaloshinskii-Moriya interaction. In \textbf{Fig.~\ref{Fig.5}(d)}, the ample strength of magnetoelectric coupling dominates the response in a saturated FM state. \textbf{Movie 5} in the Supplemental Material animates these cases \cite{Supplementary}.

To determine the size of a Skyrmion, we use the spin-plot, and count the number of magnetic spins in a Skyrmion. An example of spin-plot is shown in \textbf{Fig.~\ref{Fig.5}(e)}, which involves two Skyrmions. The number of magnetic spins contributing to these Skyrmions can be counted, which is the size of a Skyrmion. Therefore, a phase diagram parameterized by the total size of Skyrmions versus the strength of magnetoelectric coupling $g^*$ is presented in \textbf{Fig.~\ref{Fig.5}(f)}. In this figure, four kinds of regime are apparent. (1) For smaller $g^*$, the size of Skyrmions is zero due to the magnetoelectric energy being insufficient to generate Skyrmions, but the lattice forms spin spiral structure. (2) Slightly larger $g^*$ gives a mixed regime of spin spirals and Skyrmions co-existing in the lattice. (3) Certain larger magnitudes of $g^*$ generate stable Skyrmions. This region shows a non-linear decrease of the size of a Skyrmion with the increased magnetoelectric coupling. (4) For larger $g^*$, the uniform magnetization appears in this stage, due to the magnetoelectric energy dominating energy contribution in the FM system. More details are shown in the Supplemental Material \textbf{Fig. 1} \cite{Supplementary}.

\section{Conclusion}
\label{Conclusion}

This Article has shown that by using the converse magnetoelectric effect in a composite CM/FE bilayer, magnetic Bloch Skyrmions can be induced and manipulated by attaching a mobile polarized FE thin film. The quantity and quality of these Skyrmions correspond to the conditions: (1) Propagation velocity. High speed of the polarized FE layer restricts the number of Skyrmion formations. (2) Thin film size. Larger space requires more Skyrmions to minimize the local energy contribution; (3) Magnetoelectric coupling strength. The competition between the magnetoelectric interaction and the Dzyaloshinskii-Moriya interaction may occur different states in chiral-magnets. Excessive strength favors in the centrosymmetric structure, like in ferromagnetism; insufficient strength shows a spin spiral state. Only a delicate balance of the magnetoelectric coupling can offer the existence of magnetic Skyrmions.

Single-phase multiferroics have persistent coupling between the magnetic moments and the electric moments, due to their solid crystallographic structures. But, the coupling in the composite multiferroics can be easily varied by different material combinations, external stress or heat. Therefore, it opens a novel approach on magnetoelectric-induced Skyrmions in the composite bilayer. From an application viewpoint, our proposal has the potential lead to a unique technology for the future Skyrmion based memory devices.

\begin{acknowledgments}
	The authors thank X.C. Zhang and F. Xu for discussions. Z.W. gratefully acknowledges Wang Yuhua, Zhao Bingjin, Zhao Wenxia and Wang Feng for support.
\end{acknowledgments}

\bibliography{Bibliography}

\end{document}